\documentclass[sort&compress
    ,final            
  ]
  {aipproc}

\layoutstyle{6x9}

\usepackage{amsmath}
\usepackage{amssymb}
\usepackage{graphicx}
\usepackage{fancyhdr}
\usepackage{bm}

\newcommand{\cc}{cosmological constant}

\newcommand{\Cal}[1]{\ensuremath{\mathcal{#1}}}
\newcommand{\dV}{\ensuremath{\partial\Cal{V}}}
\newcommand{\LL}{Lanczos-Lovelock }
\newcommand{\D}{\ensuremath{\nabla}}
\newcommand{\Riem}[4]{\ensuremath{R^{#1 #2}_{#3 #4}}}
\newcommand{\Alt}[6]{\ensuremath{\delta^{#1 #2 ... #3}_{#4 #5
      ... #6}}} 

\newcommand{\AltC}[8]{\ensuremath{\delta^{#1 #2 #3... #4}_{#5 #6 #7
      ... #8}}}  
\newcommand{\sD}[1]{\sum_{m=1}^{K}{#1}}
\newcommand{\LDm}{\ensuremath{\Cal{L}^{(D)}_m}}
\newcommand{\eqn}[1]{Eq.\eqref{#1}}
\newcommand{\ph}[1]{\phantom{#1}}

\begin{document}

\title{Gravity as an emergent phenomenon: A conceptual description}

\classification{\texttt{04.60.-m,04.70.Dy}}

\author{T. Padmanabhan}{
  address={  IUCAA, Post Bag 4, Ganeshkhind,\\
 Pune - 411 007, India\\
email: paddy@iucaa.ernet.in}
}

\begin{abstract}
I describe several broad features of  a programme to understand gravity as an emergent,
long wavelength, phenomenon (like elasticity) and discuss one concrete framework for realizing this paradigm in the backdrop of several recent results. 
\end{abstract}

\maketitle


\section{An approach to quantum gravity}

In this article I will describe a possible approach to quantum gravity which  provides some new insights concentrating on conceptual and broader issues. Technical details can be found in the papers in ref. \cite{reviews}.

To understand the spirit behind this programme, let us begin by recalling the role played by principle of equivalence in the development of general relativity. The fact, that accelerations of two different masses in a given gravitational field are independent of their masses, was known for centuries. Einstein's genius was in realizing that this particular feature requires an explanation and then developing a theoretical framework (`gravity is curvature of spacetime') in which it emerges naturally. This is in contrast with a purely technical approach he could have taken in combining gravity with special relativity by, say, changing $\nabla^2\phi$ to $\square^2\phi$ in Poisson's equation etc. 

One can similarly visualize two different approaches while attempting to combine the principles of quantum theory and gravity. One is purely technical which hopes to exploit the `unreasonable effectiveness of mathematics' in describing Nature \cite{wigner} and works towards obtaining technical breakthroughs. The other will be to identify features of gravity that  require explanation in any quantum version and develop a paradigm which will incorporate these features in a natural fashion. Though these two approaches, technical and conceptual, need to overlap to a great extent, there\textit{ is} a  recognizable shift in the emphasis between the two approaches. My discussion will be tuned towards the latter.

The  toughest part of this approach will be to ask the right questions and identify features of gravity which needs to be incorporated in any theory of quantum gravity.  Once this is done we need to develop a mathematical structure in which these features are embedded naturally --- like the way principle of equivalence was embedded in general relativity. So I will begin by listing the possible set of features that need to incorporated in any paradigm that hopes to bring together the principles of quantum theory and gravity and then provide a (partially) working model.

{\it 1. Horizons are inevitable in the theory and they are always observer dependent.}

Principle of equivalence implies that trajectories of light will be affected by gravity. So in any theory which links gravity to spacetime dynamics, we can have nontrivial null surfaces which block information from certain class of observers. The existence of horizons is a  feature of 
any geometrical theory  of  gravity and is reasonably independent of the field equations (i.e., dynamics) of the theory.

What is more, the horizon is \textit{always} an observer dependent concept, even when it can be given a purely geometrical definition.
For example, the $r=2M$ surface in Schwarzschild geometry (which can be defined through the boundary of the causal past of $\mathcal{I}^+$)  acts as a horizon only for the class of observers who choose to stay at $r>2M$; observers falling into the black hole will have access to more information and $r=2M$ surface is not, operationally speaking, a horizon for them. 
This also means that we
should not make artificial distinctions between the Rindler horizon in flat spacetime and, say, event horizon of a black hole. 

The operational definition of an observer-dependent horizon should be causal; one should be able to decide whether a surface is a horizon or not at time $t=T$ without knowing the evolution of the spacetime geometry at $t>T$. One possible way is to choose a congruence $\mathcal{C}$ of timelike curves and define the horizon $\mathcal{H(\mathcal{C})}$ associated with $\mathcal{C}$ to be the boundary of the union of the  causal past of $\mathcal{C}$. In such an approach, any null surface will act `locally'  as a horizon for a class of observers.

\textit{2. Thermal nature of horizons cannot arise without the spacetime having a microstructure.}

In the study of ordinary solids, one can distinguish between three levels of description. At the macroscopic level,
we have the theory of elasticity which has a life of its own and can be developed purely phenomenologically.  At the other extreme,  the microscopic description of a solid will be in terms of the statistical mechanics of a lattice of atoms and their interaction. 
Interpolating between these two limits  is the thermodynamic description of a solid at finite temperature which provides a crucial window into the existence of the corpuscular substructure of solids. As Boltzmann taught us, heat is a form of motion
and \textit{we will not have the thermodynamic layer of description if  matter is a continuum all the way to the finest scale} and atoms did not exist! The mere existence of a thermodynamic layer in the description is proof enough that there are microscopic degrees of freedom. 

Move on from a solid to the spacetime. Again we should have three levels of description. The macroscopic level is the smooth spacetime continuum with a metric tensor $g_{ab}(x^i)$ and the  equations governing the metric have the same status as the phenomenological equations of elasticity. At the microscopic level, we  expect a quantum description in terms of the `atoms of spacetime' and some associated degrees of freedom $q_A$ which are still elusive. But what is crucial is the existence of an interpolating layer of thermal phenomenon associated with null surfaces in the spacetime.  Just as a solid  cannot exhibit thermal phenomenon if it does not have microstructure, horizons cannot exhibit thermal behaviour if spacetime has no microstructure.

The usual picture of treating the
metric as incorporating the dynamical degrees of freedom of the theory
is therefore not fundamental and the metric must be thought of as a
coarse grained description of the spacetime at macroscopic scales
(somewhat like the density of a solid which has no meaning at atomic  
scales). Normally we would expect the  microscopic structure of spacetime
to manifest itself only at scales comparable to Planck length (which would be analogous to the lattice spacing of a solid
\cite{zeropoint}). However, in a manner which is not fully understood,
the horizons link \cite{magglass} certain aspects of microscopic
physics with the bulk dynamics, just as thermodynamics  provides a
link between statistical mechanics and (zero temperature) dynamics of
a solid. The  reason is probably related to the fact that horizons
lead to infinite redshift, which probes \textit{virtual} high energy
processes; it is, however, difficult to establish this claim in
mathematical terms. 

{\it 3. All observers have a right to describe physics using an effective theory based only on the variables (s)he can access.}

This was, of course, the lesson from renormalization group theory. If you want to describe physics at 10 GeV you shouldn't need to know what happens at $10^{14}$ GeV in "good" theories.
In ref. \cite{adpatel} this was translated to a specific physical principle in the context of the observers having only limited access to the spacetime region.  If a class of observers perceive a horizon, they should still be able to do physics using only the variables accessible to them withough having to know what happens on the other side of the horizon. One possible way of ensuring this is to add a suitable boundary term to the action principle which will provide additional information content for observers who perceive a horizon. Mathematically,  it is convenient to treat the relevant null surface which acts as a horizon as a limit of a sequence of timelike surfaces. (For example, we can think of $r=2M$ surface as the limit of a sequence of surfaces $r=2M+\epsilon$ surfaces when $\epsilon\to0$.) The surface term in the action is also defined through a similar limiting procedure.

{\it 4. Cosmological constant problem arises due to our misunderstanding of the nature of gravity.}

In  classical general relativity, described by a Lagrangian $L=(c^3/16\pi G)(R-2\Lambda)$ with three constants $G,c,\Lambda$, there are no dimensionless constants. But when we bring in $\hbar$ we get a dimensionless number $(G\hbar/c^3)\Lambda$ which is either zero or enormously tiny.
 \textit{In either case}, it requires an explanation. (We do have some observational indications \cite{ccevidence}
suggesting it is  nonzero but the \cc\ is a problem  even if it is zero.). The trouble is that matter sector (ignoring supersymmetry, which is anyway broken at a high energy) is invariant under the shifting of the Lagrangian by a constant $L_m\to L_m+\rho$. If our theory is generally covariant, then the integration measure in the action will be $\sqrt{-g}d^Dx$ and 
gravity will break this symmetry through the coupling $\rho\sqrt{-g}d^Dx$ and will change the value of the cosmological constant. This means you cannot really solve the cosmological constant problem (that is, predict its value - zero or a tiny one) in any theory with an action principle which is (i) generally covariant (ii) uses $g_{ab}$ as dynamical variables that are varied in the action and (iii) has a matter sector that is invariant under the shift $L_m\to L_m+\rho$. This strongly suggests that we shouldn't look for an action principle in which
the metric is varied to obtain the equations of motion for gravity.

{\it 5. Gravity is an emergent phenomenon like elasticity and the field equations should be derivable from an alternative paradigm.}

I will call this the Sakharov paradigm \cite{elastic} and it is fairly obvious from all that  has been said before. But one can add two more concrete ingredients.
First,
there is a deep connection between the dynamical equations governing the metric and the thermodynamics of horizons. An explicit example  was provided in ref.
\cite{paddy2}, in the case of spherically symmetric horizons 
 in which it was shown that, Einstein's equations can be interpreted as a 
thermodynamic relation $TdS=dE+PdV$ arising out of virtual
radial displacements of the horizon. Further work showed that this result is valid in \textit{all} the cases for which explicit computation can be carried out --- like in the
Friedmann models 
\cite{rongencai} as well as for rotating and time dependent horizons
in Einstein's theory \cite{dawood-sudipta-tp}. 

Second,
the Einstein-Hilbert Lagrangian has the structure $L_{EH}\propto R\sim (\partial g)^2+ {\partial^2g}$.
In the usual approach the surface term arising from  $L_{sur}\propto \partial^2g$ has to be ignored or canceled to get Einstein's equations from  $L_{bulk}\propto (\partial g)^2$.  
But there is a 
peculiar (unexplained) relationship between $L_{bulk}$ and $L_{sur}$:
\begin{equation}
    \sqrt{-g}L_{sur}=-\partial_a\left(g_{ij}
\frac{\partial \sqrt{-g}L_{bulk}}{\partial(\partial_ag_{ij})}\right)
\end{equation}
This shows that the gravitational action is `holographic' with the same information being coded in both the bulk and surface terms. This suggests that either one of them  is sufficient.  Of course, it is well known that varying $g_{ab}$ in $L_{\rm bulk}$ leads to the standard field  equations. More remarkable is the fact that one can also obtain Einstein's equations from an action principle which  uses only the surface term and the  virtual displacements of horizons \cite{paris}
\textit{without} treating the metric as a dynamical variable. 
Since the surface term has the thermodynamic interpretation as the entropy of horizons, this establishes a direct connection between spacetime dynamics and horizon thermodynamics.
This also suggests that surface displacements (encoded in suitable vector fields) can be used to obtain the 
dynamical equations without the metric having to be varied in an action principle.

{\it 6. Einstein's theory is just a low energy effective theory; the thermodynamic description should provide clues as to how to obtain the corrections to this theory.}

The  above description has shifted the emphasis from Einstein's general relativity --- described by a particular action principle, field equations, etc. --- to a broader picture of spacetime thermodynamics of horizons. This suggests that we should be led to wider class of theories of which Einstein gravity is just a special case.

There is already some work indicating the nature of higher order corrections to the theory. Recent work has shown that \textit{all the  thermodynamic features described above extend far beyond Einstein's theory.}
 The connection between field equations and the thermodynamic relation $TdS=dE+PdV$ 
 is not restricted to
Einstein's theory (GR) alone, but is in fact true for the
case of the generalized, higher derivative \LL gravitational theory in
$D$ dimensions as well \cite{aseem-sudipta}. The same is true  for the holographic structure of the action functional \cite{ayan}: the \LL action has the same structure and --- again --- the entropy of the horizons is related to the surface term of the action. \textit{These results show that the thermodynamic description is far more general than just Einstein's theory} and occurs in a wide class of theories in which the metric determines the structure of the light cones and null surfaces exist blocking the information.


\section{Key aspects of the new paradigm}

How do we translate these features into a mathematical formalism for describing gravity? 
Recall that the local inertial frames played a crucial role in developing the \textit{kinematics} of general relativity and (almost) fixes the manner in which gravity affects other matter. To obtain a similar guiding principle regarding the \textit{dynamics} of gravity we need to proceed one step further and introduce the notion of a local rindler frame (LRF). Around (\textit{not} `at') any event $\mathcal{P}$, we can introduce a Riemann normal coordinates in which the metric has the form $g_{ab}=\eta_{ab}+\mathcal{O}(Rx^2)$ where $R$ symbolically stands for the components of the curvature tensor. We now boost the coordinates with a (sufficiently large) uniform acceleration $g$ along an arbitrary direction to introduce a LRF around $\mathcal{P}$.
 The null surface $X=T$ of the local inertial frame will act as the future horizon in the LRF and will be at a distance $1/g$. For $g\gg R^{1/2}$, there will be a local neighborhood in which the physics will be that in a Rindler frame. (If we introduce an analytic extension to complex time in LRF, the the null surfaces $X=\pm T$ will be mapped to the origin and the region beyond the null surfaces will disappear. The local nature is obvious in the Euclidean sector. Noncompact region near the horizons $X=\pm T$ in the $XT$ plane gets mapped to a compact region around the origin in the Euclidean
 plane.)  The ideas described in item 2 in the previous section suggest that the physics will now be governed by a pure surface term on the Rindler horizon. It is possible to determine the form of this surface term from general considerations. If we now demand that the action should not receive contributions for radial displacements of the horizons, defined in a particular manner, one can obtain --- at the lowest order --- the equations
 \begin{equation}
(G_{ab}-\kappa T_{ab})\xi^a\xi^b=0
\label{nullvec}
\end{equation} 
where $\xi^a$ is a null vector. (I have skipped several details which are elaborated in the previous publications). Demanding the validity of \eqn{nullvec} in all LRF then leads to Einstein's theory with the cosmological constant emerging as an integration constant. Note that \eqn{nullvec} is invariant under the constant shift of matter Lagrangian, making gravity immune to bulk cosmological constant.

This result leads to several further  conclusions:
\begin{itemize}
\item 
The surface terms in the actions can lead to sensible equations of motion when the variational principle is suitably designed. This is consistent with the `holographic' nature of gravitational theories mentioned earlier. (In fact, in the Riemann normal coordinates, the bulk term of the Einstein-Hilbert Lagrangian vanishes and only the surface term survives to give $L_{EH}=L_{bulk}+L_{sur}=L_{sur}\propto R$.)
\item
We are not varying the metric to get the equations of motion and, in fact, when you only have a surface term you can't get anything sensible by that route. The variational principle used here is very peculiar and essentially allows you to extract a $TdS$ term from the surface term (which is actually the entropy)
and arrives at something equivalent to $TdS=dE+pdV$. This \textit{one} equation has the contents equivalent to \textit{ten} Einstein's equations when we demand that it holds in \textit{every} LRF.
\item
The procedure can be generalized to get the \LL theory in a similar manner from a surface term. The higher order terms can be interpreted as corrections to Einstein's equations. As mentioned earlier, the thermodynamic route is more general than Einstein's theory and does not use metric as a dynamical variable.

\end{itemize}

In obtaining the above results, one  treats the null surfaces as the limit of a sequence of, say, timelike surfaces. The displacements of the horizon used in the action principle will be in the direction normal to the surfaces. All these suggest that one may be able to obtain a more formal description of the theory in terms of normalized vector fields in spacetime
without sticking to null vectors and surfaces. Such a generalization is anyway needed for the theory to make sense in the Euclidean sector in which vectors cannot be separated as  null, spacelike or timelike. I will now describe one such model which is unreasonably successful.

To set the stage, let us suppose there are certain microscopic --- as yet unknown --- degrees of freedom $q_A$, analogous to the atoms in the case of solids, described by some microscopic action functional $A_{micro}[q_A]$. In the case of a solid, the relevant long-wavelength elastic dynamics is captured by  the \textit{displacement vector field}
which occurs in the equation $x^a\to x^a+\xi^a(x)$. In the case of spacetime, we no longer want to use metric as a dynamical variable; so we  need to introduce some other degrees of freedom, analogous to $\xi^a$ in the case of elasticity, and an effective action functional based on it.  Normally, varying an action functional with respect certain degrees of freedom will lead to equations of motion determining \textit{those} degrees of freedom.  But we now make an unusual demand that varying our  action principle with respect to some (non-metric) degrees of freedom should lead to an equation of motion \textit{determining the background metric} which remains non-dynamical.

Based on the role expected to be played by surfaces in spacetime, we shall take the relevant degrees of freedom to be the normalized vector fields $n_i(x)$ in the spacetime \cite{aseementropy} with a  norm that  is fixed at every event but might vary from event to event: (i.e., $n_in^i\equiv\epsilon(x)$ with $\epsilon(x)$ being a fixed function which takes the values $0,\pm1$ at each event.) 
Just as the displacement vector $\xi^a$ captures the macro-description in case of solids, the  normalized vectors (e.g., normals to surfaces) capture the essential macro-description in case of gravity in terms of an effective action $S[n^a]$. More formally, we expect the coarse graining of microscopic degrees of freedom to lead to an effective action in the  long wavelength limit:
\begin{equation}
\sum_{q_A}\exp (-A_{micro}[q_A])\longrightarrow \exp(-S[n^a])
\label{microtomac}
\end{equation} 
To proceed further we need to determine the nature of $S[n^a]$. The general form of $S[n^a]$ in such an effective description, at the quadratic order, will be:
\begin{equation}
S[n^a]=\int_\Cal{V}{d^Dx\sqrt{-g}}
    \left(4P_{ab}^{\ph{a}\ph{b}cd} \D_cn^a\D_dn^b - 
    T_{ab}n^an^b\right) \,,
\label{ent-func-2}
\end{equation}
where $P_{ab}^{\ph{a}\ph{b}cd}$ and $T_{ab}$ are two tensors and the signs, notation etc.  are chosen with hindsight. (We will see that $T_{ab}$ can be identified with the matter stress-tensor.)
The full action for gravity plus matter will be taken to be $S_{tot}=S[n^a]+S_{matt}$ with:
\begin{equation}
S_{tot}=\int_\Cal{V}{d^Dx\sqrt{-g}}
    \left(4P_{ab}^{\ph{a}\ph{b}cd} \D_cn^a\D_dn^b - 
    T_{ab}n^an^b\right)+\int_\Cal{V}{d^Dx\sqrt{-g}} \mathcal{L}_{matter}
\end{equation}  
with an important extra prescription: Since the gravitational sector is related to spacetime microstructure, we must \textit{first} vary the $n^a$ and \textit{then} vary the matter degrees of freedom. In the language of path integrals, we should  integrate out the gravitational degrees of freedom $n^a$ first and use the resulting action for the matter sector. 
 
We next address the crucial  difference 
between the dynamics in gravity and say, elasticity,  which we mentioned earlier. In the case of solids, one will write a similar functional for thermodynamic potentials in terms of the displacement vector $\xi^a$ and extremising it will
lead to an equation  \textit{which determines}
$\xi^a$. In the case of spacetime, we expect the variational principle to hold for   all vectors $n^a$ with constant norm and lead to a condition on  the \textit{background
metric.} Obviously, the action functional in \eqn{ent-func-2} must be rather special to accomplish this and one need to impose two restrictions on the coefficients $P_{ab}^{\ph{a}\ph{b}cd}$ and $T_{ab}$ to achieve this.  First, the tensor $P_{abcd}$ should
have the algebraic symmetries similar to the Riemann tensor $R_{abcd}$
of the $D$-dimensional spacetime. Second, we need:
\begin{equation}
\D_{a}P^{abcd}=0=\D_{a}T^{ab}\,.
\label{ent-func-1}
\end{equation}
In a
complete theory, the explicit form of $P^{abcd}$ will be determined by the
long wavelength limit of the microscopic theory just as the elastic
constants can --- in principle --- be determined from the microscopic
theory of the lattice. In the absence of such a theory, we can take a cue from
the renormalization group theory and expand $P^{abcd}$
in powers of  derivatives of
the metric \cite{paris,aseementropy}. That is, we expect,
\begin{equation}
P^{abcd} (g_{ij},R_{ijkl}) = c_1\,\overset{(1)}{P}{}^{abcd} (g_{ij}) +
c_2\, \overset{(2)}{P}{}^{abcd} (g_{ij},R_{ijkl})  
+ \cdots \,,
\label{derexp}
\end{equation} 
where $c_1, c_2, \cdots$ are coupling constants and the successive terms progressively probe smaller and smaller scales.  The lowest order
term must clearly depend only on the metric with no derivatives. The next
term depends (in addition to metric) linearly on curvature tensor and the next one will be quadratic in curvature etc. It can be shown that  the m-th order term
which satisfies our constraints is \textit{unique} and is given by
\begin{equation}
\overset{(m)}{P}{}_{ab}^{\ph{a}\ph{b}cd}\propto
\AltC{c}{d}{a_3}{a_{2m}}{a}{b}{b_3}{b_{2m}}
\Riem{b_3}{b_4}{a_3}{a_3} \cdots
\Riem{b_{2m-1}}{b_{2m}}{a_{2m-1}}{a_{2m}} 
 =
\frac{\partial\LDm}{\partial R^{ab}_{{\ph{ab}cd}}}\,. 
\label{LL03}
\end{equation}
where $\AltC{c}{d}{a_3}{a_{2m}}{a}{b}{b_3}{b_{2m}}$ is the alternating tensor and the last equality shows that it
can be  expressed
as a derivative of the m th order \LL Lagrangian \cite{paris,lovelock}, given by
\begin{equation}
\Cal{L}^{(D)} = \sD{c_m\LDm}\,~;~\Cal{L}^{(D)}_m = \frac{1}{16\pi}
2^{-m} \Alt{a_1}{a_2}{a_{2m}}{b_1}{b_2}{b_{2m}}
\Riem{b_1}{b_2}{a_1}{a_2} \Riem{b_{2m-1}}{b_{2m}}{a_{2m-1}}{a_{2m}}
\,,  
\label{LL222}
\end{equation}
where the $c_m$ are arbitrary constants and \LDm\ is the $m$-th
order \LL term and we assume
$D\geq2K+1$.
The lowest order term (which leads to Einstein's theory) is
\begin{equation}
\overset{(1)}{P}{}^{ab}_{cd}=\frac{1}{16\pi}
\frac{1}{2} \delta^{a_1a_2}_{b_1b_2} =\frac{1}{32\pi}
(\delta^a_c \delta^b_d-\delta^a_d \delta^b_c)
  \,.
\label{pforeh}
\end{equation} 
while the first order term (which gives the  Gauss-Bonnet correction) is:
\begin{equation}
\overset{(2)}{P}{}^{ab}_{cd}= \frac{1}{16\pi}
\frac{1}{2} \delta^{a_1a_2a_3a_4}_{b_1\,b_2\,b_3\,b_4}
R^{b_3b_4}_{a_3a_4} =\frac{1}{8\pi} \left(R^{ab}_{cd} -
         G^a_c\delta^b_d+ G^b_c \delta^a_d +  R^a_d \delta^b_c -
         R^b_d \delta^a_c\right) 
\label{ping}
\end{equation} 
All higher orders terms are obtained in a similar manner.

In our paradigm based on \eqn{microtomac}, the field equations for gravity arises from extremising $S$ with respect to
variations of the  vector field $n^a$, with the constraint $\delta (n_an^a)=0$, and demanding that the
resulting condition holds for \textit{all normalized vector fields}.  
One can show that this leads to the field equations
\begin{equation}
16\pi\left[ P_{b}^{\ph{b}ijk}R^{a}_{\ph{a}ijk}-\frac{1}{2}\delta^a_b\LDm\right]=
 8\pi T{}_b^a +\Lambda\delta^a_b   
\label{ent-func-71}
\end{equation}
These are identical to the field equations for \LL gravity with a cosmological constant arising as an undetermined integration constant.  To the lowest order, when we use \eqn{pforeh} for $P_{b}^{\ph{b}ijk}$, the \eqn{ent-func-71} reproduces Einstein's theory. More generally, we get Einstein's equations with
higher order corrections which are to be interpreted as emerging  
from the derivative expansion of the action functional as we probe smaller and smaller scales. 
Remarkably enough, we can derive not only Einstein's theory but even \LL theory from a dual description in terms on the normalized vectors in spacetime, \textit{without varying $g_{ab}$ in an action functional!}

The crucial feature of the coupling between matter and gravity through $T_{ab}n^an^b$ is that, under the shift $T_{ab}\to T_{ab}+\rho_0g_{ab}$ the $\rho_0$  term in the action in \eqn{ent-func-2} decouples from $n^a$ and becomes irrelevant:
\begin{equation}
\int_\Cal{V}{d^Dx\sqrt{-g}}T_{ab}n^an^b \to 
\int_\Cal{V}{d^Dx\sqrt{-g}} T_{ab}n^an^b +
\int_\Cal{V}{d^Dx\sqrt{-g}}\epsilon\rho_0
\end{equation} 
Since $\epsilon=n_an^a$ is not varied when $n_a$ is varied there is no coupling between $\rho_0$ and the dynamical variables $n_a$ and the theory is invariant under the shift  $T_{ab}\to T_{ab}+\rho_0g_{ab}$.
 We see that the condition $n_an^a=$ constant on the  dynamical variables have led to a `gauge freedom' which allows an arbitrary integration constant to appear in the theory which can absorb the bulk cosmological constant. This is important but that is another story (see the first paper in ref.\cite{reviews} and ref.\cite{vacfluc}).  
   
To gain a bit more insight into what is going on, let us consider the on-shell value of the action functional.  
Manipulating
the covariant derivatives in \eqn{ent-func-2} and using the  field equation \eqn{ent-func-71} we can show that the only dependence on $n_a$ is through a surface term. 
Explicitly, the \textit{on-shell} value of this surface term is given by:
\begin{equation}
S=4\int_{\dV}{d^{D-1}x\sqrt{h}\,k_a\left(P^{abcd} n_c\D_b n_d\right)}
\longrightarrow  -\frac{1}{8\pi}\int_{\dV}d^{D-1}x\sqrt{h}\,k_i\left(n^iK+a^i\right)
\label{on-shell-2}
\end{equation}
where $k^i$ is the normal to the boundary and  we have manipulated a few indices using the symmetries of
$P^{abcd}$. The  second expression, after the arrow, is the result for
general relativity. Note that the integrand has the
familiar structure of $k_i( n^iK+a^i)$ where $a^i= n^b\D_b n^i$ is
the acceleration associated with the vector field $ n^a$ and
$K\equiv -\D_b n^b$ is the trace of extrinsic
curvature in the standard context. 
If we restrict to a series of surfaces foliating the spacetime with $n_i$ representing their unit normals and take the boundary to be one of them, we can identify $k_i$ with $n_i$; then $a_in^i=0$ and the surface term is just
\begin{equation}
S|_{\rm on-shell}=\mp\frac{1}{8\pi}\int_{\dV}d^{D-1}x\sqrt{h} K
\label{ygh}
\end{equation} 
which is the York-Gibbons-Hawking boundary term in general relativity
\cite{gh} if we normalise $\epsilon$ to $\pm1$ depending on the nature of the surface.

It is now obvious that this term in the on-shell action will lead to the entropy of the horizons (which will be 1/4 per unit transverse area) in the case of general relativity.  In fact, the result is far more general. Even in the case of of a more general $P^{ab}_{cd}$ it can be shown that the on-shell value of the action reduces to \cite{aseementropy} the entropy of the horizons. The general expression is:
\begin{equation}
S|_{\Cal{H}} = \sD{4\pi m c_m \int_{\Cal{H}}{d^{D-2}x_{\perp} 
  \sqrt{\sigma}\Cal{L}^{(D-2)}_{(m-1)}}} 
 =\frac{1}{4}[{\rm Area}]_\perp +{\rm corrections}     
\label{ent-limit-2}
\end{equation} 
where $x_{\perp}$ denotes the transverse coordinates on the horizon \Cal{H},
$\sigma$ is the determinant of the intrinsic metric on \Cal{H} and we
have restored a summation over $m$ thereby giving the result for the
most general \LL case obtained as a sum of individual \LL lagrangians.  The expression in \eqn{ent-limit-2} \emph{is
  precisely the entropy of a general Killing horizon in \LL gravity}
based on the general prescription given by Wald and others
\cite{noether} and computed by several authors.  
Further, in any spacetime, 
if we take a local Rindler frame around  any event 
we will obtain an entropy for the locally defined Rindler horizon. In the case of GR, this entropy per unit transverse area is just 1/4 as expected. 

This result shows that, in the semiclassical limit in which the action can possibly be related to entropy, we reproduce the conventional entropy which scales as the area in Einstein's theory. Since the entropy counts the relevant degrees of freedom, this shows that the degrees of freedom which survives and contributes in the long wave length limit  scales as the area. The quantum fluctuations in these degrees of freedom can then lead to the correct, observed, value of the \cc.
(The last aspect can be made more quantitative \cite{vacfluc} but we will not do it here.)

Our action principle is somewhat peculiar compared to the usual action principles in the sense that we have varied $n_a$ and demanded that the resulting equations hold for \textit{all} vector fields of constant norm. So the action principle actually stands for an infinite number of action principles, one for each vector field of constant norm! This class of \textit{all} $n^i$ allows an effective, coarse grained, description of some (unknown) aspects of spacetime micro physics. 

\section{Conclusions}

These results suggest an attractive conceptual framework for realizing the paradigm mentioned in the beginning. The key idea is to introduce local Rindler observers around each event in a spacetime and demand that they have a right to describe physics using variables which are accessible to them. This requires using an action principle which is a pure boundary term and studying its variation under  the normal displacements of the horizon. Equivalently, one can think of all the normalized vector fields in the spacetime as describing the virtual displacements of surfaces and write down an effective action in terms of these vector fields. The variation of either action --- in a manner which is quite different  from the standard variational principles --- leads to a general set of field equations. Einstein's theory emerges as the lowest order term with \LL type higher order corrections. As a bonus, one gets a handle on \cc\ problem, which I have described elsewhere.
The fact that one can push these ideas consistently this far is remarkable.

\end{document}